\documentclass[prl,twocolumn,superscriptaddress]{revtex4}
\usepackage{amssymb,amsmath,amsthm,graphicx}
\usepackage{latexsym}
\usepackage{bm}
\usepackage[]{hyperref} 

\def\be{\begin{equation}}
\def\ee{\end{equation}}
\def\beq{\begin{eqnarray}}
\def\eeq{\end{eqnarray}}

\pdfoutput=1

\begin{document}

\def\lsim{\mathrel{\rlap{\lower4pt\hbox{\hskip1pt$\sim$}}
    \raise1pt\hbox{$<$}}}
\def\gsim{\mathrel{\rlap{\lower4pt\hbox{\hskip1pt$\sim$}}
    \raise1pt\hbox{$>$}}}
\def\be{\begin{equation}}
\def\ee{\end{equation}}
\def\bea{\begin{eqnarray}}
\def\eea{\end{eqnarray}}
\newcommand{\der}[2]{\frac{\partial{#1}}{\partial{#2}}}
\newcommand{\dder}[2]{\partial{}^2 #1 \over {\partial{#2}}^2}
\newcommand{\dderf}[3]{\partial{}^2 #1 \over {\partial{#2} \partial{#3}}}
\newcommand{\eq}[1]{Eq.~(\ref{eq:#1})}
\newcommand{\dd}{\mathrm{d}}

\title{Coleman-de Luccia reconsidered: a subtlety of gravity and the thin wall approximation}

\author{Keith Copsey}
\email{kcopsey@perimeterinstitute.ca}
\affiliation{ Perimeter Institute for Theoretical Physics, Waterloo, Ontario N2L 2Y5, Canada\\
\&  Department of Physics \& Astronomy, University of Waterloo, Waterloo, Ontario N2L 3G1, Canada}

\begin{abstract}
I point out that the usual Coleman-de Luccia analysis of tunnelling via instantons between perturbatively stable minima using the thin-wall approximation misses one of the effects of gravitational backreaction on the on-shell action and hence the decay rate.  Once this oversight is corrected, one finds these decay rates are much larger than has been generally appreciated; including the effects of gravity potentials involving barriers which are relatively high and not overly wide result in decays which are quite rapid instead of slow.  In the light of these results, it is no longer clear that one should believe string theory predicts a wide class of cosmologically long-lived metastable de Sitter vacua.
\end{abstract}

\maketitle


The use of instantons to describe decays between perturbatively stable minima was pionereed by Coleman and collaborators in a series of classic papers \cite{Colemantunnelling}.  This analysis was extended to include the effects of gravity by Coleman and de Luccia \cite{CdL}.  The point of this letter is to point out, in as concise a fashion as possible, that the particular method of evaluating the on-shell action used in \cite{CdL} ignores one of the effects of gravitational backreaction and ends up dropping a term which ought to be retained.   This error ends up being far more significant than one might have naively thought and correcting this oversight one finds dramatically enhanced decay rates.   Aside from this technical point, the calculation proceeds in a fashion very much along the lines of \cite{CdL}, although for the sake of brevity, various detailed justifications and discussions of related issues will be presented elsewhere \cite{CopseyTunnelling}.

Given an instanton interpolating between two points near perturbatively stable minima, as any solution starting out precisely at a minima stays there indefinitely, the tunnelling rate may be calculated via
\be \label{dcyrate}
\Gamma = A e^{-B}
\ee
where $B$ is the difference between the on-shell Euclidean action for the instanton minus the on-shell action  $S_b$ of the original ``background'' solution (e.g. a false vacuum)
\be
B = S_E - S_b
\ee
While the calculation of the prefactor $A$ including backreaction remains an open problem, in the absence of special symmetries (e.g. fermionic zero modes) $A$ is presumably at least as large as the volume of the instanton in Planck units.   Provided $B$ is sufficiently large, then $A$ is merely a logarithmic correction and only important if one is calculating $B$ to quite high accuracy.  

Begininning with the Euclidean action in $d$ dimensions for gravity and a minimally coupled scalar field
\be \label{origact}
S_E = -\kappa \int \sqrt{g} \Big(R - \frac{1}{2} (\nabla \phi)^2 - V(\phi)\Big)
\ee
where $\kappa = (16 \pi G_d)^{-1}$ and I have taken the usual (i.e. Gibbons-Hawking-Perry \cite{GHPactn}) overall sign of the action so that the rate of black hole production in de Sitter is exponentially suppressed rather than enhanced \cite{Bousso:1996wy, Bousso:1996wz}.  Taking the trace of Einstein's equation one finds
\be
R = \frac{1}{2} (\nabla \phi)^2 + \frac{d}{d-2} V(\phi)
\ee
and then the on-shell action may be written as
\be \label{act1}
S_E = -\frac{2 \kappa}{d-2} \int \sqrt{g} \, V(\phi)
\ee
Note the effect of gravity has had two rather remarkable effects--the on-shell action no longer has any gradient terms and higher values of the potential, presuming the backreaction effects on the metric do not dominate, make the action more negative not more positive.  The later effect is most easily achieved if the increase in potential occurs in a small region, as the metric and hence determinant are necessarily continuous.  On the other hand, if the potential is nearly uniform the backreaction effects on the metric are important, as in, for example, de Sitter space where the action is proportional to $V^{1-d/2}(\phi)$ and a larger potential results in a less negative action.

One might reasonably expect the most symmetric non-trivial instantons to dominate the decay rate, so let us consider a metric
\be \label{metansatz}
ds^2 = d\tau^2 + \rho^2(\tau) d\Omega_{d-1}
\ee
where $d\Omega$ is the metric on the unit $(d-1)$-sphere and $\phi(\tau)$. 
With these assumptions the Einstein and field equations are equivalent to
\be \label{Ein1}
{\rho'}^2 = 1 + \frac{\rho^2}{(d - 1) (d - 2)} \Big(\frac{1}{2} {\phi'}^2 - V(\phi) \Big)
\ee
\be \label{Ein2}
\frac{\rho''}{\rho} = -\frac{{\phi'}^2}{2 (d - 1)} - \frac{V(\phi)}{(d - 1) (d - 2)}
\ee
and
\be \label{Feqn}
\phi'' + (d - 1) \frac{\rho'}{\rho} \phi' = V'(\phi)
\ee
For $\phi$ beginning precisely in some potential minima,  the unique regular solution to the above is $\phi(\tau) = \phi_0$, for some constant $\phi_0$, and
\be \label{rhosol1}
\rho = \frac{1}{\omega} \sin \omega \tau
\ee
where
\be \label{omega0def}
\omega^2 = \frac{V(\phi_0)}{(d-1) (d-2)}
\ee
provided one takes the obvious limit $(\rho = \tau)$ in the case $V(\phi_0) = 0$.  In the case of $V(\phi_0) > 0$ (i.e. de Sitter space), (\ref{metansatz}) describes a $d$-sphere.  More generally, in the asymptotically de Sitter case (i.e. positive false vacuum potential), starting at $\rho = 0$ when $\tau = 0$, eventually $\rho''$ (\ref{Ein2}) always becomes negative and $\rho$ goes through a second zero at some finite $\tau$, resulting in a topologically spherical instanton.  In the asymptotically flat or asymptotically AdS case the instanton is topologically a disk with a single value of $\tau$ where $\rho$ vanishes and $\rho \rightarrow \infty$ as $\tau \rightarrow \infty$.  Continuing a polar angle in the $(d-1)$-sphere (\ref{metansatz}) along a zero extrinsic curvature surface one obtains the Lorentzian solutions  produced by the tunnelling \cite{Turokopeninflat}.  Given the above symmetry ansatz  the on-shell action (\ref{act1}) becomes 
\be \label{act2}
S_E = -\frac{2 \kappa \Omega_{d-1}}{d-2} \int_{0}^{\tau_f} d\tau \rho^{d-1} V(\phi)
\ee
where $\tau$ is chosen so that $\rho(0) = 0$ in all cases and in the asymptotically dS case $\rho(\tau_f) = 0$ as well, while in the asymptotically flat or asymptotically AdS case $\tau_f \rightarrow \infty$.

While there does not appear to be any straightforward method to analytically solve the Einstein equations for a generic potential even given the above simple ansatz, Coleman and de Luccia \cite{CdL} pointed out that a reasonably broad class of instantons could be analyzed if one assumed that $\rho(\tau)$ were approximately constant when $\phi$ is not near one of the minima.   Somewhat more precisely, this thin-wall approximation will hold when the amount that $\rho$ changes as $\phi$ crosses the barrier between the minima is small compared to the value of $\rho$.  In turn, this occurs if $\phi$ starts rather close to a potential minima and only leaves that region once $\rho$ becomes relatively large.  In terms of the potential this situation will occur, broadly speaking, when the potential barrier is relatively high and not overly wide.  In particular, if the width of the barrier is small in Planck units, the thin-wall approximation will hold provided the height of the barrier is large compared to the difference between the minima.   It is also possible to show the approximation is valid for a variety of potentials of interest where the barrier is of order a Planck width, including the model KKLT potential \cite{KKLT}.

Given the thin-wall approximation, the action may be split into three pieces as a near-true vacuum region, a near-false vacuum region, and a wall interpolating between these two regions:
\beq \label{act5}
S_E &=& -\frac{2 \kappa \Omega_{d-1}}{d-2}\Bigg[  \underbrace{\int_{0}^{\tau_0} d\tau \rho^{d-1} V(\phi)}_{I_0} + \underbrace{\int_{\tau_0}^{\tau_1} d\tau \rho^{d-1} V(\phi)}_{I_1} \nonumber \\ 
&+& \underbrace{\int_{\tau_1}^{\tau_f} d\tau \rho^{d-1} V(\phi)}_{I_2}  \Bigg]
\eeq
where $I_0$ corresponds to the nearly true vacuum region where one may approximate $V(\phi) = V_0$ and $\rho$ is given by (\ref{rhosol1}) with $V(\phi_0) \rightarrow V_0$, $I_1$ the action for the wall where, to a good approximation, $\rho = \rho_0$ for some constant $\rho_0$, and $I_2$ for the nearly false vacuum region where $V(\phi) = V_2$ and
\be
\rho = \frac{1}{\omega_2} \sin \omega_2 (t - \delta)
\ee
with $\omega_2$ defined as in (\ref{omega0def}) and $\delta$ chosen so that $\rho(\tau_0) = \rho(\tau_1) = \rho_0$.  Since in the wall region one may approximate $\rho = \rho_0$,
\be
I_1 =  \rho_0^{d-1} \int_{\tau_0}^{\tau_1} d \tau V(\phi) = \frac{V_2}{\omega_2^d} (\omega_2 \rho_0)^{d-1} V_a
\ee
where the (dimensionless) wall tension is defined
\be
V_a = \omega_2 \int_{\tau_0}^{\tau_1} d\tau \frac{V(\phi)}{V_2} = \omega_2 \int_{\phi(\tau_0)}^{\phi(\tau_1)} \frac{d\phi}{\phi'} \frac{V(\phi)}{V_2}
\ee
At leading order in the thin-wall approximation $V_a$ depends only on the potential and is independent of $\rho_0$.  Aside from a neighborhood of the barrier maximum, where a slightly more subtle argument is required, the last statement is just the observation of Coleman and de Luccia \cite{CdL} that $\rho$ is relatively large away from the minima of the potential and as a result the friction term in the field equation (\ref{Feqn}) is negligable and $\phi'' \approx V'(\phi)$.  As a result of the above observations, in the thin-wall approximation the only free parameter in the on-shell action is $\rho_0$; all the remaining functions depend only on the potential.   Since the action is by definition extremized by any solution, then it must be true that $\partial_{\rho_0} S_E = 0$ and this allows one to solve for $\rho_0$.

In the work of Coleman and de Luccia, as well as the succeeding work by Parke \cite{ParkeBubbles}, rather than extremize $S_E$ they propose to extremize $B$, breaking up $S_b$ in an analogous fashion to (\ref{act5}) 
\beq \label{act6}
S_b &=& -\frac{2 \kappa \Omega_{d-1}}{d-2}\Bigg[  \underbrace{\int_{0}^{\tau_0} d\tau \hat{\rho}^{d-1} \hat{V}(\phi)}_{\hat{I}_0} + \underbrace{\int_{\tau_0}^{\tau_1} d\tau \hat{\rho}^{d-1} \hat{V}(\phi)}_{\hat{I}_1}  \nonumber \\
&+& \underbrace{\int_{\tau_1}^{\hat{\tau}_f} d\tau \hat{\rho}^{d-1} \hat{V}(\phi)}_{\hat{I}_2}  \Bigg]
\eeq
where the false vacuum background potential $\hat{V}(\phi)=V_2$ and
\be
\hat{\rho} = \frac{1}{\omega_2} \sin \omega_2 \tau
\ee
and then calculate $B$ by subtracting the integrands of $I_0 - \hat{I}_0$ and so forth 
\be
B =  -\frac{2 \kappa \Omega_{d-1}}{d-2}\Bigg[ \int_0^{\tau_0} d \tau [ \rho^{d-1} V(\phi) - \hat{\rho}^{d-1} \hat{V}(\phi)]  + \ldots \Bigg]
\ee
Strictly speaking, \cite{CdL} and \cite{ParkeBubbles} use a slightly different form of the on-shell action than the above, obtained by imposing the ansatz (\ref{metansatz}) on the original action (\ref{origact}) and performing an integrating by parts to remove the second order derivatives; in the asymptotically flat or asymptotically AdS case their approach may be subject to challenge as they simply disregard the resulting (divergent) surface term but in the asymptotically de Sitter case there is no such surface term (the instanton is compact without boundary) and these two actions are strictly equivalent.   The above form, however, will be more useful later.

The problem with this approach is that generically $\tau_f \neq \hat{\tau}_f$; the evolution of $\rho$ depends upon the potential in question (\ref{Ein2}) and the amount of time, namely $\tau_f$, it takes for $\rho$ to go from zero, reach a maximum, and go back to zero in the asymptotically de Sitter case or the amount of time for $\rho$ to get to a given large value, for the other asymptotics, is a function of the potential.   In particular, as $\phi$ travels over a barrier the increasing potential makes $\rho''$ more negative, both directly and through the effects of an increasing $\phi'^2$, than it would have been had $\phi$ stayed in the false vacuum background.   If one tries to subtract integrands then one will generically find a mismatch in the ranges of $\tau$ between $S_E$ and $S_b$ and hence one ends up with a leftover piece of one of the integrals that can not be paired with the other.  Specifically, the assertion in \cite{CdL} and \cite{ParkeBubbles} in the above notation is that $I_2 = \hat{I}_2$.  While it is true the potentials in these two integrals match, $\rho(\tau) \neq \hat{\rho}(\tau)$ for $\tau > \tau_1$ unless $\tau_f = \hat{\tau}_f= \pi/{\omega_2}$;  given an approximately constant potential and regularity $\rho(\tau)$ has a unique solution and these different endpoints imply $\rho(\tau_1) \neq \hat{\rho}(\tau_1)$.  Since, by assumption, in the thin-wall approximation $\rho$ is relatively large at the wall this error can be substantial.  One could avoid this mismatch by aligning the zero of the instanton as $\phi$ approaches the false vacuum, instead of the true vacuum, with a zero of the background solution but then the mismatch will occur as $\phi$ approaches the true vacuum.   Regardless of how one chooses the zero of $\tau$, the fact that the amount of Euclidean time the instanton lasts differs from the background solution means one can not subtract integrands without a mismatch.

It is, however, easy to avoid this error by merely extremizing $S_E(\rho_0)$ or by extremizing $B(\rho_0)$ while treating $S_b$ as simply a constant rather than trying to subtract integrands.   In the asymptotically de Sitter case (decays from a positive false vacuum) a bit of straightforward algebra shows, except for transitions to sufficiently negative potential minima, the condition $\partial_{\rho_0} S_E = 0$ specifies $\rho_0$ uniquely and the on-shell action of the instanton $S_E$ is actually less than that of the background.  This implies the decay rate, instead of being exponentially suppressed, is exponentially enhanced.  For example, for decays from de Sitter minima $V(\phi) = V_2 > 0$ to flat space in four dimensions one finds
\be \label{rho04d}
\rho_0 = \frac{ 3 V_a}{\omega_2 \sqrt{1 + 9 V_a^2}}
\ee
and
\be
B = S_E - S_b = -\frac{72  \pi^2 \kappa}{V_2}  \frac{\Big(1-\sqrt{1+9 V_a^2}\Big)^2}{3 \sqrt{1 + 9 V_a^2}}
\ee
and hence $B < 0$ for any positive $V_a$.  The fact that this rate can be exponentially enhanced is entirely due to the fact, as discussed above, that in the wall region due to gravitational effects a larger potential (i.e. the barrier) makes the action more negative, not more positive.  It may not be immediately obvious that this effect dominates over the fact that the instanton has a smaller volume than the background solution but this result is easily confirmed by either performing the above calculation or, if one prefers, making a quick order of magnitude estimate \cite{CopseyTunnelling} for the case of large tension.

Note that, unlike the Coleman-de Luccia result, $\rho_0$ (\ref{rho04d}) is a monotonically increasing function of the wall tension and in the large tension limit approaches the false vacuum scale; the previous results asserted that large tension instantons caused the entire universe, not just a region that could be described within a causal patch, to tunnel  to a small volume and hence, as has been noted before \cite{EvadS}, conflicted with causality.   One might be surprised that the decay rate grows as $V_a$ increases.  For systems where gravity is not important this type of behavior would be difficult to understand but, as discussed above, once gravity is included this should not be unexpected.  More broadly speaking, gravity rewards configurations with large energy densities confined in small regions, as is well known dynamically (gravity makes matter clump) and thermodynamically (negative specific heats). 

The possible exception to this exponentially enhanced decay from positive false vacuum minima $V_F$ in the thin-wall approximation comes if the true vacuum potential $V_T$ is negative and of a magnitude larger than the false vacuum potential.  There remains  one root $\rho_0$ extremizing the action and for which $S_E < S_b$ for all choices of the potential parameters.  However, there is also a region in the parameter space $(V_T/V_F, V_a)$ where two additional roots $\rho_0$ which extremize $S_E$, satisfy the obvious criterion for sensibility, and correspond to actions less negative than $S_b$.  Naively, the exponentially enhanced instanton would dominate the decay, but in this case a more thorough investigation into these modes, including checking the  sign of the second variational derivative of the action, is in order.  The existence of multiple solutions also appears to contradict an argument of Jensen and Steinhardt \cite{SteinhardtJensenI} regarding the uniqueness of instanton solutions and deserves investigation.

The discussion of the decay of asymptotically flat solutions is, in principle, complicated by the need to add appropriate surface terms to the action to ensure a well-defined variational principle as well as deal with any cutoff surface ambiguities.  To the best of my knowledge, neither of these issues has been investigated in detail for the cosmological context but at least a naive estimate suggests while these concerns may well be significant in the asymptotically AdS case in the asymptotically flat case neither is important and the above bulk action as it stands is complete.   Assuming this to be so, then the only difference with the above approximations is that one should approximate $\rho = \tau - \delta$ for $\tau \geq \tau_1$,
\be
\rho = \frac{1}{\omega_0} \sinh(\omega_0 \tau)
\ee
for $0 \leq \tau \leq \tau_0$ where
\be
\omega_0^2 = - \frac{V_0}{(d-1) (d-2)}
\ee
and the previous definition of tension breaks down ($V_2 \rightarrow 0$) so it is useful to define the closely related
\be
\tilde{V}_a = \omega_0 \int_{\tau_0}^{\tau_1} d \tau \frac{V(\phi)}{(-V_0)}
\ee
Focusing, for the sake of simplicity, on the $d = 4$ decay from flat space to negative potential minima $V(\phi) = V_0$ one finds a $\rho_0$ that extremizes $S_E$ if and only if $ 0< \tilde{V}_a < 1/3$ at
\be
\rho_0 = \frac{3 \tilde{V}_a}{\omega_0 \sqrt{1 - 9 \tilde{V}_a^2}}
\ee
and then
\be
B =  S_E = -\frac{72  \pi^2 \kappa}{\vert V_0 \vert}  \frac{\Bigg(1-\sqrt{1-9 \tilde{V}_a^2}\Bigg)^2}{3 \sqrt{1 - 9 \tilde{V}_a^2}}
\ee
One again one finds exponentially enhanced decay rates.  The fact that there is a bound on $\tilde{V}_a$ is the fact, pointed out by Coleman-de Luccia \cite{CdL} and a variety of authors since, that it becomes impossible to write down a zero energy bubble configuration for sufficiently large wall tension since large volumes in anti-de Sitter space scale only as areas.   

The decay of an asymptotically anti-de Sitter solution would appear to contradict not only the AdS/CFT correspondence \cite{AdSCFT, BanksTasi} but a variety of positive energy theorems in anti-de Sitter space \cite{PosEnergyAdS} showing the only zero energy solution is pure AdS.   Nevertheless, there does not appear to be any obstruction to constructing instantons that would naively describe such a decay; it is not hard to numerically construct such examples even without making any thin-wall approximation.   It would be interesting to understand how to resolve this tension.

Rather than the above, the thin-wall approximation has sometimes in the literature been taken to mean one may treat the wall as a $\delta$-function surface and merely use the junction conditions to calculate the instanton.   Strictly speaking, such a solution is a naked singularity (the square of the Riemann tensor will diverge at such a surface) and if one has in mind modelling a thin but regular configuration one requires some argument that the described configuration is the limit of some regular family of solutions.   In the asymptotically flat case this is usually a perfectly reasonable way to model a brane, so it may be a bit surprising that at least in the asymptotically de Sitter case this last criterion actually fails.   While leaving the details to \cite{CopseyTunnelling}, one can show explicitly that if one tried to take the limit of an infinitely thin but finite tension barrier and treat $\rho$ as a constant across the barrier, $\phi$ necessarily acquires a nonzero velocity as it crosses the wall region and as a result the instanton is singular.  Taking the limit of an increasingly thin barrier of a finite regular system one can show that in order for the thin-wall approximation to hold, in the sense described above, one is forced to take a small tension and small bubble limit as the barrier becomes arbitrarily thin.

The usual argument that string theory allows a vast number of metastable vacua rests heavily on the clever KKLT construction \cite{KKLT} describing a relatively tall barrier separating a de Sitter minima from flat space.   While relatively tall barriers appear to be inevitably necessary to ensure a long-lived vacua with relatively low cosmological constant (one needs to ensure reheating, let alone astronomical phenomena or even particle accelerator collisions, do not provide enough energy to travel directly over the barrier), unless the barrier is relatively wide as well it will give rise to an instanton well-described by the thin-wall approximation and hence, once one includes the effects of gravity, decay rapidly.   On the other hand, very wide barriers would appear to generically give rise to low mass scalar fields which are objectionable not just on phenomenological grounds but also from the viewpoint of radiative stability--it appears difficult to understand why such scalars should not have a mass at least of the order of the supersymmetry-breaking scale.  Constructing long-lived de Sitter vacua then appears to require either finding special points in the landscape that in all of the hundreds if not thousands of directions in field space avoid these problems and give rise to effective potentials that are not well-described by the thin-wall approximation or by finding a way to avoid the usual decompactification problem and construct a truly stable de Sitter minima.

\vskip .5cm
\centerline{\bf Acknowledgements}
\vskip .2cm
It is a pleasure to thank N. Turok, L. Boyle, N. Afshordi, A. Brown, and M. Johnson for useful discussions.  This work was supported in part by the Natural Sciences \& Engineering Research
Council of Canada.   Research at Perimeter Institute is supported by the Government of Canada through Industry Canada and by the Province of Ontario through the Ministry of Research \& Innovation.


\end{document}